\documentclass[prl,twocolumn,superscriptaddress]{revtex4-1}
\usepackage{braket}
\usepackage{upgreek}
\usepackage{array}
\usepackage{amssymb}
\usepackage{amsmath}
\usepackage{amsfonts}
\usepackage{bbm}
\usepackage{graphicx}
\usepackage{float} 
\usepackage{subfigure}
\usepackage{mathrsfs}
\usepackage{graphics,graphicx,epsfig,bm,amsmath,amsthm,amssymb}
\usepackage{bm}
\usepackage{bbm}
\usepackage{longtable}
\usepackage{multirow}
\usepackage{array}
\usepackage{color}
\usepackage{diagbox}

\usepackage[usenames,dvipsnames]{xcolor}

\usepackage{float}
\usepackage[a4paper,colorlinks=true,
linkcolor=blue,citecolor=blue,
pdfauthor={ },
pdftitle={ },
pdfsubject={ },
pdfkeywords={ }]{hyperref}

\begin{document}

\title{Experimental observation of Dirac exceptional point}

\affiliation{CAS Key Laboratory of Microscale Magnetic Resonance and School of Physical Sciences, University of Science and Technology of China, Hefei 230026, China}
\affiliation{Hefei National Laboratory, University of Science and Technology of China, Hefei 230088, China}
\affiliation{Anhui Province Key Laboratory of Scientific Instrument Development and Application, University of Science and Technology of China, Hefei 230026, China}
\affiliation{Institute of Quantum Sensing and School of Physics, Zhejiang University, Hangzhou 310027, China}

\author{Yang Wu}
\author{Dongfanghao Zhu}
\affiliation{CAS Key Laboratory of Microscale Magnetic Resonance and School of Physical Sciences, University of Science and Technology of China, Hefei 230026, China}
\affiliation{Anhui Province Key Laboratory of Scientific Instrument Development and Application, University of Science and Technology of China, Hefei 230026, China}

\author{Yunhan Wang}
\affiliation{CAS Key Laboratory of Microscale Magnetic Resonance and School of Physical Sciences, University of Science and Technology of China, Hefei 230026, China}
\affiliation{Hefei National Laboratory, University of Science and Technology of China, Hefei 230088, China}
\affiliation{Anhui Province Key Laboratory of Scientific Instrument Development and Application, University of Science and Technology of China, Hefei 230026, China}


\author{Xing Rong}
\email{xrong@ustc.edu.cn}
\affiliation{CAS Key Laboratory of Microscale Magnetic Resonance and School of Physical Sciences, University of Science and Technology of China, Hefei 230026, China}
\affiliation{Hefei National Laboratory, University of Science and Technology of China, Hefei 230088, China}
\affiliation{Anhui Province Key Laboratory of Scientific Instrument Development and Application, University of Science and Technology of China, Hefei 230026, China}

\author{Jiangfeng Du}
\email{djf@ustc.edu.cn}
\affiliation{CAS Key Laboratory of Microscale Magnetic Resonance and School of Physical Sciences, University of Science and Technology of China, Hefei 230026, China}
\affiliation{Hefei National Laboratory, University of Science and Technology of China, Hefei 230088, China}
\affiliation{Anhui Province Key Laboratory of Scientific Instrument Development and Application, University of Science and Technology of China, Hefei 230026, China}
\affiliation{Institute of Quantum Sensing and School of Physics, Zhejiang University, Hangzhou 310027, China}

\begin{abstract}                                   
The energy level degeneracies, also known as exceptional points (EPs), are crucial for comprehending emerging phenomena in materials and enabling innovative functionalities for devices. 
Since EPs were proposed over half a century age, only two types of EPs have been experimentally discovered, revealing intriguing phases of materials such as Dirac and Weyl semimetals. 
These discoveries have showcased numerous exotic topological properties and novel applications, such as unidirectional energy transfer.
Here we report the observation of a novel type of EP, named the Dirac EP, utilizing a nitrogen-vacancy center in diamond.
Two of the eigenvalues are measured to be degenerate at the Dirac EP and remain real in its vicinity.
This exotic band topology associated with the Dirac EP enables the preservation of the symmetry when passing through, and makes it possible to achieve adiabatic evolution in non-Hermitian systems.
We examined the degeneracy between the two eigenstates by quantum state tomography, confirming that the degenerate point is a Dirac EP rather than a Hermitian degeneracy.
Our research of the distinct type of EP contributes a fresh perspective on dynamics in non-Hermitian systems and is potentially valuable for applications in quantum control in non-Hermitian systems and the study of the topological properties of EP.
\end{abstract}

\maketitle

\begin{figure*}[http]
	\centering
	\includegraphics[width=2\columnwidth]{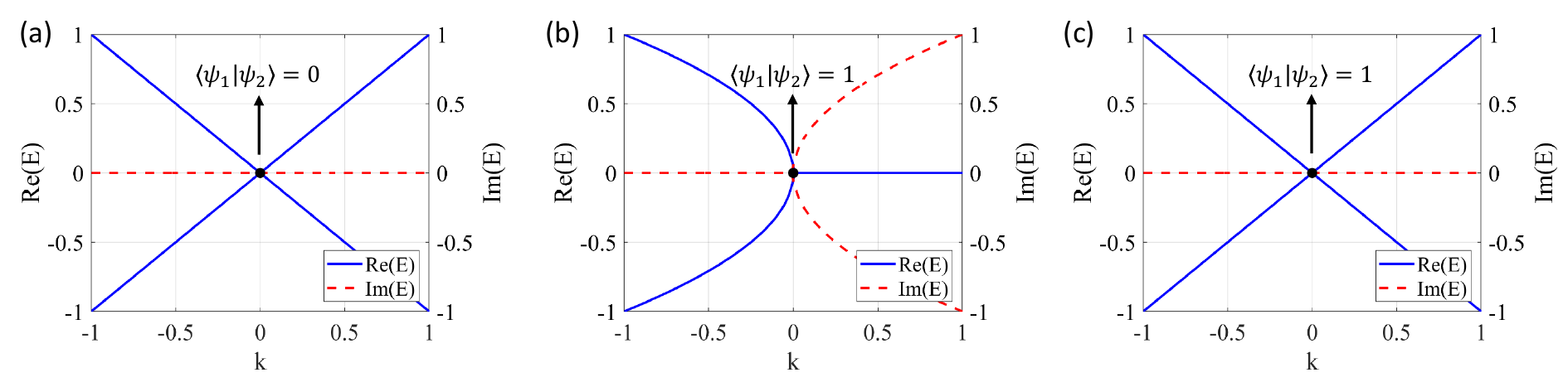}
	\caption{Degeneracy of Hermitian system and typical EP and Dirac EP of non-Hermitian system. (a-c) The real (blue lines) and imaginary (red dashed lines) parts of the Hamiltonian hosting different type of degeneracies. (a) Near the Hermitian degeneracy, the eigenvalues are always real and vary linearly with the parameter $k$. The eigenstates $\ket{\psi_1}$ and $\ket{\psi_2}$ are orthogonal at the Hermitian degeneracy. (b) The eigenvalues change from real to imaginary at typical EP and exhibit a square root dependence on the parameters. The eigenstates coalesce at typical EP. (c) In vicinity of the Dirac EP, the eigenvalues are real and have a linear relationship with the parameters. The eigenstates are degenerate at the Dirac EP.
	}
	\label{Fig1}
\end{figure*}

\textit{Introduction}- Exceptional points (EPs) were introduced in mathematics to characterize the energy level degeneracy of a linear operator over half a century ago~\cite{Book_Kato}.
Since EPs were proposed, only two types of EPs have been experimentally discovered.
One type is the degeneracies in the Hermitian system, such as diabolical points~\cite{LSA_Yang,NNano_Polimeno}.
The eigenvalues are always real and vary linearly with the parameter near the Hermitian degeneracy.  
At the Hermitian degeneracy, the eigenvalues are degenerate, but the eigenstates are orthogonal (Fig.~\ref{Fig1}a).
This type of EP has been extensively studied, revealing intriguing phases of materials such as Dirac and Weyl semimetals~\cite{RMP_Yarkony}.
The other is the typical EPs introduced by non-Hermitian physics~\cite{PRL_Bender,PRL_Yao,PRX_Kawabata,PRX_Gong,Science_Zhou,NP_Xiao,PRL_Hassan}, where both eigenvalues and eigenstates are degenerate (Fig.~\ref{Fig1}b).
Typical EPs in non-Hermitian systems are usually accompanied by changes in symmetries, like parity-time ($\mathcal{PT}$), anti-$\mathcal{PT}$, or particle-hole symmetry. 
A non-Hermitian system undergoes a corresponding breaking or restoration of symmetry across the EP. 
This occurrence is often marked by significant alternations in the energy spectrum, shifting between real and imaginary values.
Investigations of typical EPs not only have deepened the understanding of topological physics, such as novel non-Hermitian topological phases~\cite{RMP_Bergholtz,PRL_Hu,PRL_Shen,PRL_Kawabata}, exceptional nodal topologies~\cite{PRB_Stalhammar,NRP_Ding} and topological state control~\cite{PRL_Hassan}, but also have given rise to a variety of potential applications, such as single-mode lasing~\cite{Science_Feng,Science_Hodaei} and unidirectional invisibility~\cite{NatPhoto_Chang,NP_Peng,PRL_Lin,Nature_Xu}.
Moreover, the fractional exponent response of eigenvalues to parameter perturbations such as a square and cubic root relationship near a second order and third order EP, implies potentially enhanced sensitivity~\cite{Nature_Chen,Nature_Hodaei,NP_Park,SA_Mao}.

\begin{figure}[http]
	\centering
	\includegraphics[width=1\columnwidth]{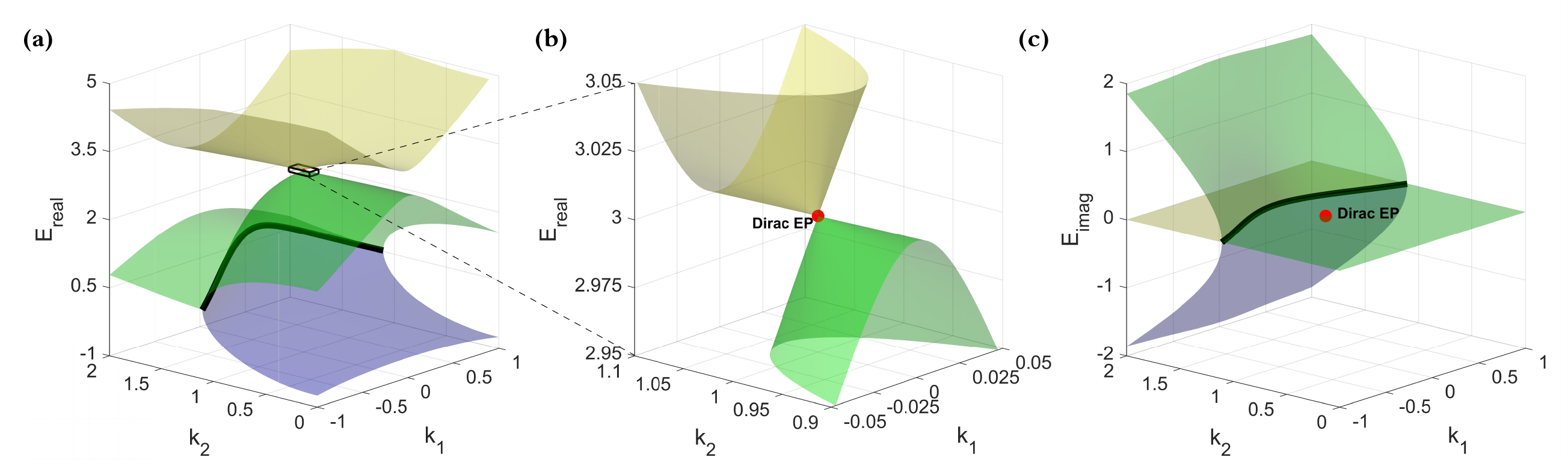}
	\caption{The eigenvalues structure and the Dirac exceptional points of the non-Hermitian Hamiltonian $H(k_1,k_2)$.  (a,b) The coloured surfaces are the eigenvalue sheets of $H(k_1,k_2)$ for the real (a) and imaginary (c) parts. Yellow, green and purple sheets correspond to eigenvalues $E_1$, $E_2$ and $E_3$, respectively. The Dirac EP marked by the red dot is located at the $k_1=0$ and $k_2=1$, where the eigenvalues $E_1$ and $E_2$ degenerate. (b) The eigenvalues structure near the Dirac EP of the non-Hermitian Hamiltonian showing the Dirac cone. The imaginary parts of the three eigenvalues near Dirac EP are zero. An exceptional line consisting entirely of typical EPs is represented by the black line.
	}
	\label{Fig2}
\end{figure}

Here we report the experimental observation of a novel type of EPs, named the Dirac EPs~\cite{PRL_Rivero}, utilizing a nitrogen-vacancy (NV) center in diamond.
The simultaneous degeneracy of two of the eigenvalues and eigenstates clearly distinguishes the Dirac EP from the Hermitian degeneracy (Fig.~\ref{Fig1}c).
The eigenvalues measured near the EP are all real which maintain the parity-time symmetry when passing through the Dirac EP.
Due to the real eigenvalues around the Dirac EP, non-adiabatic transitions that occur when encircling typical EPs no longer appear.
Moreover, the two eigenvalues connected at the Dirac EP exhibit a linear and conical dispersion with the change of the parameters.
The multitude of exotic properties exhibited by the Dirac EP, in contrast to the typical EPs, deepens our understanding of mechanism for dynamics in non-Hermitian systems and holds potential for the investigation of mode switch without dissipation and adiabatic evolution in non-Hermitian systems.
\textit{Theoretical model}- To observe the Dirac EP, we construct a $\mathcal{PT}$-symmetric non-Hermitian Hamiltonian in the tight binding model as follows~\cite{PRL_Rivero}:
\begin{equation}
	\begin{aligned}
		H_{\mathrm{tb}}&=\sum_{m\in \mathbb{Z}}(m+k_1)^2\ket{m}\bra{m}+\frac{1-k_2}{2}\ket{m}\bra{m+1} \\
		&+\frac{1+k_2}{2}\ket{m}\bra{m-1},		
	\end{aligned}
	\end{equation}
where $k_1$ represents momentum and $k_2$ characterizes the magnitude of the non-reciprocity of the coupling between adjacent lattices. 
By truncating the block of $m=-1, 0$ and $1$ in the tight binding model~\cite{PRL_Rivero}, the $3\times3$ non-Hermitian Hamiltonian can be obtained as follows:
\begin{equation}
	H(k_1,k_2)=3S_z^2+2k_1S_z+\sqrt{2}(S_x-ik_2S_y),
	\label{Hamiltonian}
\end{equation}
where $S_x$, $S_y$ and $S_z$ are spin-1 operators.
The Hamiltonian in Eq.~\ref{Hamiltonian} naturally has a Dirac EP in its parameter space, without introducing any high-order terms of parameters, which is a must for constructing Dirac EPs in two band models~\cite{PRB_Rivero}.
The positions of EPs are determined by the condition that the characteristic polynomial $P(E)\equiv \mathrm{det}[H(k_1,k_2)-E]=f_3E^3+f_2E^2+f_1E+f_0$ has multiple roots, where $f_3=-1$, $f_2=6$, $f_1 =4k_1^2-2k_2^2-7$, $f_0 = 6(k_2^2-1)$ (See Supplementary Materials, section 2 for the detailed derivation). 
As shown in Fig.~\ref{Fig2}, an exceptional line consisting entirely of typical EPs emerges as the black line.
The Dirac EP is located at $k_1=0$, $k_2=\pm1$.
There exists a real-imaginary transition of the eigenvalues when crossing a typical EP, whereas the imaginary parts of the three eigenvalues are all zero in vicinity of the Dirac EP (Fig.~\ref{Fig2}c).
Therefore, the three eigenvalues are all real numbers near the Dirac EP.
The real parts of the two eigenvalues are also degenerate at the Dirac EP (Fig.~\ref{Fig2}a).
Unlike the square root relationship commonly observed near typical EPs, the eigenvalues exhibit a linear relationship with parameter variations in vicinity of the Dirac EP.
For any direction in the two-dimensional space, the eigenvalues satisfy the relationship $E_{1,2}=3+2(-\sin\theta\pm\sqrt{\sin^2\theta+9\cos^2\theta})\Delta k/3+O[(\Delta k)^2]$, where $\Delta k_1=\cos\theta \Delta k$, $\Delta k_2=\sin\theta \Delta k$, $\theta\in[0,2\pi]$ and $\Delta k>0$. 
Therefore, the linear relationship is satisfied in all directions provided that the parameters are very close to the Dirac EP, presenting the Dirac cone (Fig.~\ref{Fig2}b).

\textit{Experimental realizations}- The nitrogen-vacancy (NV) center, which is an atomic-scale defect in diamond, was utilized to observe the Dirac EP (See Supplementary Materials, section 1 for the experimental setup).
Dirac EP is characterized by the eigenvalue and eigenstate information of the non-Hermitian Hamiltonian $H(k_1,k_2)$, which can be obtained by the evolution of the quantum state $\ket{\psi(t)}$ under the non-Hermitian Hamiltonian.
Taking the electron spin in NV center as the system and the nuclear spin as the ancilla qubit, the evolution under non-Hermitian Hamiltonian $H(k_1,k_2)$ can be realized based on the dilation method~\cite{PRL_Gunther,Science_Wu,PRL_Liu,NatNano_Wu}. 
The evolution under non-Hermitian Hamiltonian is achieved in a subspace of the $6\times6$ dilated Hermitian Hamiltonian $H_{\rm tot}$.
The detailed procedure is as follows.
Firstly, the coupled system is prepared to the initial state $\ket{\psi(0)}_e\ket{-}_n + [\eta_0\ket{\psi(0)}_e]\ket{+}_n$, where subscripts $e$ and $n$ label the electron and nuclear spin states, respectively.
Here, $\eta_0$ is a factor properly chosen for the convenience of experimental realization, and $|\pm\rangle_n$ are the eigenstates of the Pauli matrix $\sigma_y$.
Secondly, the dilated Hamiltonian $H_{\rm tot}$ is constructed for the non-Hermitian Hamiltonian $H(k_1,k_2)$ as
\begin{equation}
	H_{\rm tot}(t)=\Gamma(t)\otimes|1\rangle_n~_n\langle1|+\Lambda(t)\otimes|0\rangle_n~_n\langle0|.
\end{equation}
Here, $\Gamma(t)$ and $\Lambda(t)$ are Hermitian operators in the following form,
\begin{equation}
	\begin{aligned}
	\Gamma(t)=\left[\begin{array}{ccc}
		d_1(t) & a_1(t) & c_1(t)\\
		a_1^*(t) & d_2(t) & b_1(t)\\
		c_1^*(t) & b_1^*(t) & d_3(t)
	\end{array}\right],\\
	\Lambda(t)=\left[\begin{array}{ccc}
		d_4(t) & a_2(t) & c_2(t)\\
		a_2^*(t) & d_5(t) & b_2(t)\\
		c_2^*(t) & b_2^*(t) & d_6(t)
	\end{array}\right],
\end{aligned}
	\label{Htot_para}
\end{equation}
where the parameters $a_i(t)$, $b_i(t)$, $c_i(t)~(i=1,2)$ and $d_j(t)~(j=1,...,6)$ are determined by the non-Hermitian Hamiltonian $H(k_1,k_2)$ according to the dilation method (See Supplementary Materials, section 5 for the derivation and the detailed expressions).
Thirdly, the dilated Hamiltonian $H_{\rm tot}$ can be realized by applying microwave and electric field control pulses in the NV center system. 
The initial state evolves to $\ket{\psi(t)}_e\ket{-}_n+[\eta(t)\ket{\psi(t)}_e]\ket{+}_n$ under the Hamiltonian $H_{\rm tot}$, where $\eta(t)$ is a properly chosen time-dependent operator. 
The effective Hamiltonian governing the evolution of $\ket{\psi(t)}_e$ in the $\ket{-}_n$ subspace corresponds to $H(k_1,k_2)$.
The diagonal elements $d_j(t)$ of $H_{\rm tot}(t)$ were realized by choosing an appropriate interaction picture .
The off-diagonal elements $a_1(t)$, $b_1(t)$, $a_2(t)$ and $b_2(t)$ of $H_{\rm tot}(t)$ were realized by four selective microwave pulses.
The off-diagonal elements $c_1(t)$ and $c_2(t)$ in $H_{\rm tot}(t)$ were implemented by alternating current electric field pulses.
The time-dependent amplitudes and phases of the microwave and electric field pulses were appropriately set according to the off-diagonal elements in $H_{\rm tot}(t)$.
Two examples of the amplitudes and phases of the microwave and electric field pulses are shown in Supplementary Materials, section 5.
Finally, the state evolving under $H(k_1,k_2)$ is obtained by postselecting the $\ket{-}_n$ subspace, leading to the determination of the eigenvalues and eigenstates of $H(k_1,k_2)$.

The characterization of the Dirac EP requires information about the eigenvalues and eigenstates of the non-Hermitian Hamiltonian.
The eigenvalues of the non-Hermitian Hamiltonian are encoded in the population of eigenstates depending on the given model (See Supplementary Materials, section 6 for the detailed derivation).
The eigenstates of an NH Hamiltonian $H(k_1,k_2)$ can be obtained from the steady states under the evolution of the NH Hamiltonian $g(H)$ ($H$ and $g(H)$ have the same eigenstates), where $g(x)$ is an analytic function of $x$.
It is worth noting that near the Dirac EP, the eigenvalues are all real.
By selecting $g(H)$ as $iH(k_1,k_2)$, $-iH(k_1,k_2)$ and $1/i(H(k_1,k_2)-cI)$, where the parameter $c$ is chosen to be closest to one of the eigenvalues, the three eigenstates can be achieved by evolving sufficiently long under NH Hamiltonian $g(H)$.
Then the eigenvalues can be obtained by measuring the populations of the eigenstates under different basis. 
Furthermore, the eigenstates can be fully characterized through quantum state tomography. 

The entire experimental procedure can be divided into three parts: the state preparation, evolution under the dilated Hamiltonian and the populations measurement. 
First, the initial state is polarized to $|0\rangle_e|1\rangle_n$ by optical pumping under the static magnetic field of 501 Gauss.
After polarization, the initial state of the coupled system is prepared in the form $|0\rangle_e(|-\rangle_n+\eta_0|+\rangle_n)$ by single-qubit rotation on the nuclear spin. 
Following the nuclear spin rotation, an appropriate operation on the electron spin prepares the coupled system into state $|\psi_{\rm ini}\rangle_e(|-\rangle_n+\eta_0|+\rangle_n)$ to ensure a sufficiently high population in the subspace during readout. 
Second, the dilated Hamiltonian $H_{\mathrm{tot}}(t)$ is implemented using microwave and electric field pulses, whose amplitudes, frequencies and phases are determined based on parameters in Eq.~\ref{Htot_para}.
After a sufficiently long evolution time, the state of the coupled system becomes $|\psi_i\rangle_e|-\rangle_n+[\eta(t)|\psi_i\rangle_e]|+\rangle_n$, where $|\psi_i\rangle_e$ is the $i^{th}$ eigenstate of the non-Hermitian Hamiltonian and $\eta(t)$ is an operator. 
Finally, the measurement basis is transformed by operations on the electron and nuclear spin before readout (See Supplementary Materials, section 6 for the detailed derivation). By renormalization in the subspace $|-\rangle_n$, measurements of populations in different bases and quantum state tomography of the three eigenstates are realized.

\begin{figure}[http]
	\centering
	\includegraphics[width=1\columnwidth]{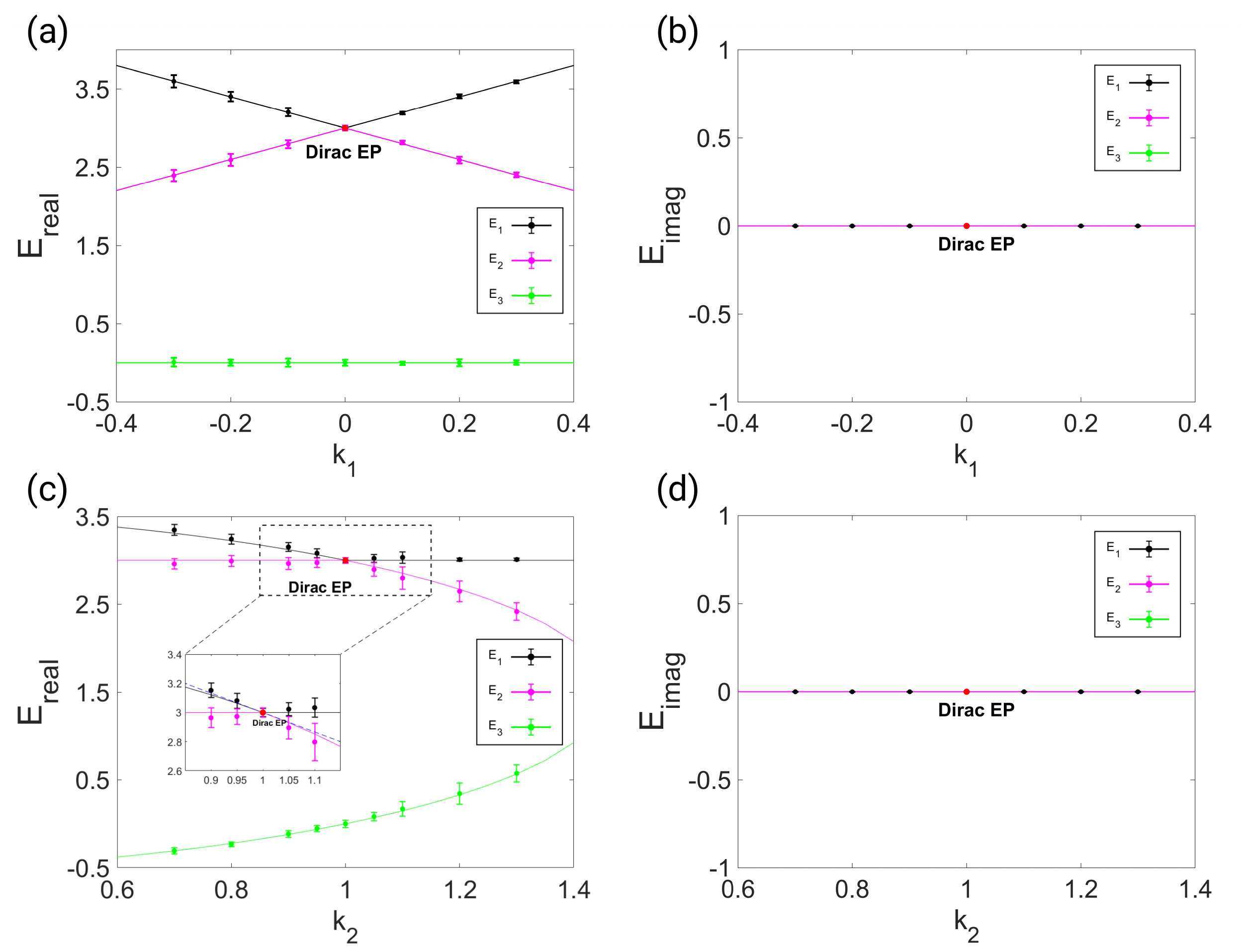}
	\caption{Observation of the Dirac EP. (a-d) The real (a,c) and imaginary (b,d) parts of the three eigenvalues of the non-Hermitian Hamiltonian. Black, red and green points and lines correspond to eigenvalues $E_1$, $E_2$ and $E_3$, respectively. When $k_2$ is set to 1, the change of eigenvalues with parameter $k_1$ confirms that the Dirac EP is located at $k_1=0$ (a,b). By fixing $k_1$ to 0, the Dirac EP can also be confirmed by observing the change of the eigenvalues with another parameter $k_2$ (c,d). The black, red and green dots with error bars are the experimental results of the three eigenvalues. The lines of the corresponding colours are theoretical predictions.
	}
	\label{Fig3}
\end{figure}

\textit{Experimental results}- Figure~\ref{Fig3} depicts the pure real eigenvalues and linear dispersion in vicinity of the Dirac EP. 
We demonstrate the eigenvalues varying with respect to two parameters separately.
When $k_2$ is set to 1, the eigenvalue $E_3$ is always zero as $k_1$ varies as shown in Fig.~\ref{Fig3}a and \ref{Fig3}b.
The eigenvalues $E_1$ and $E_2$ degenerate at the Dirac EP point when $k_1=0$.
The real parts of the eigenvalues exhibit a linear response to the parameter $k_1$, while the imaginary parts remain zero, indicating real eigenvalues in the vicinity of the Dirac EP (Fig.~\ref{Fig3}b).
A similar situation arises when $k_1$ is set to 0 and $k_2$ is varied.
The imaginary parts of the eigenvalues remain constant at 0 near the Dirac EP as shown in Fig.~\ref{Fig3}d.
Such entirely real eigenvalues ensure that the eigenstates of the non-Hermitian Hamiltonian are always eigenstates of the $\mathcal{PT}$ operator when passing through the Dirac EP, indicating the absent of the spontaneous $\mathcal{PT}$ symmetry breaking~\cite{PRB_Rivero}.
This is in stark contrast to what occurs at typical EPs, where the eigenvalues transit from real to complex when passing through them. 
Fig.~\ref{Fig3}a and \ref{Fig3}c show that the real parts of the eigenvalues degenerate at the Dirac EP. 
They also illustrate the linear relationship between the eigenvalue and the parameter when the parameter is very close to the Dirac EP.
When $k_2=1$, the eigenvalues have the forms that $E_1=3+2|k_1|$, $E_2=3-2|k_1|$, and $E_3=0$, showing a linear response to the parameter (Fig.~\ref{Fig3}a).
When $k_1$ is fixed at 0, the eigenvalues vary with $k_2$ as $E_1=\max((3+\sqrt{17-8k_2^2})/2,3)$, $E_2=\min((3+\sqrt{17-8k_2^2})/2,3)$ and  $E_3=(3-\sqrt{17-8k_2^2})/2$.
Near the Dirac EP with $k_2<1$, $E_1$ exhibits a linear variation with $\Delta k_2=k_2-1$ as $E_1=3-4\Delta k_2/3+O[(\Delta k_2)^2]$, while the value of $E_2$ remains constant at 3.
For the case of $k_2>1$, $E_1=3$ and $E_2=3-4\Delta k_2/3+O[(\Delta k_2)^2]$.
It can be observed in Fig.~\ref{Fig3}c that when $\Delta k_2$ is small, the variation of $E_1$ with $k_2$ conforms well to the linear relationship. 
However, when $\Delta k_2$ is large, influenced by higher-order terms of $\Delta k_2$, the variation of $E_1$ with $k_2$ gradually deviates from linearity.
The eigenvalues are linear in two perpendicular direction $k_1$ and $k_2$ when the parameter is very close to the Dirac EP.
These results are consistent with the dispersion relation shown in Fig.~\ref{Fig2}b.
Thus, the energy spectrum near the Dirac EP forms a structure akin to a Dirac cone.

\begin{figure*}[http]
	\vspace{0em}
	\centering
	\includegraphics[width=1.6\columnwidth]{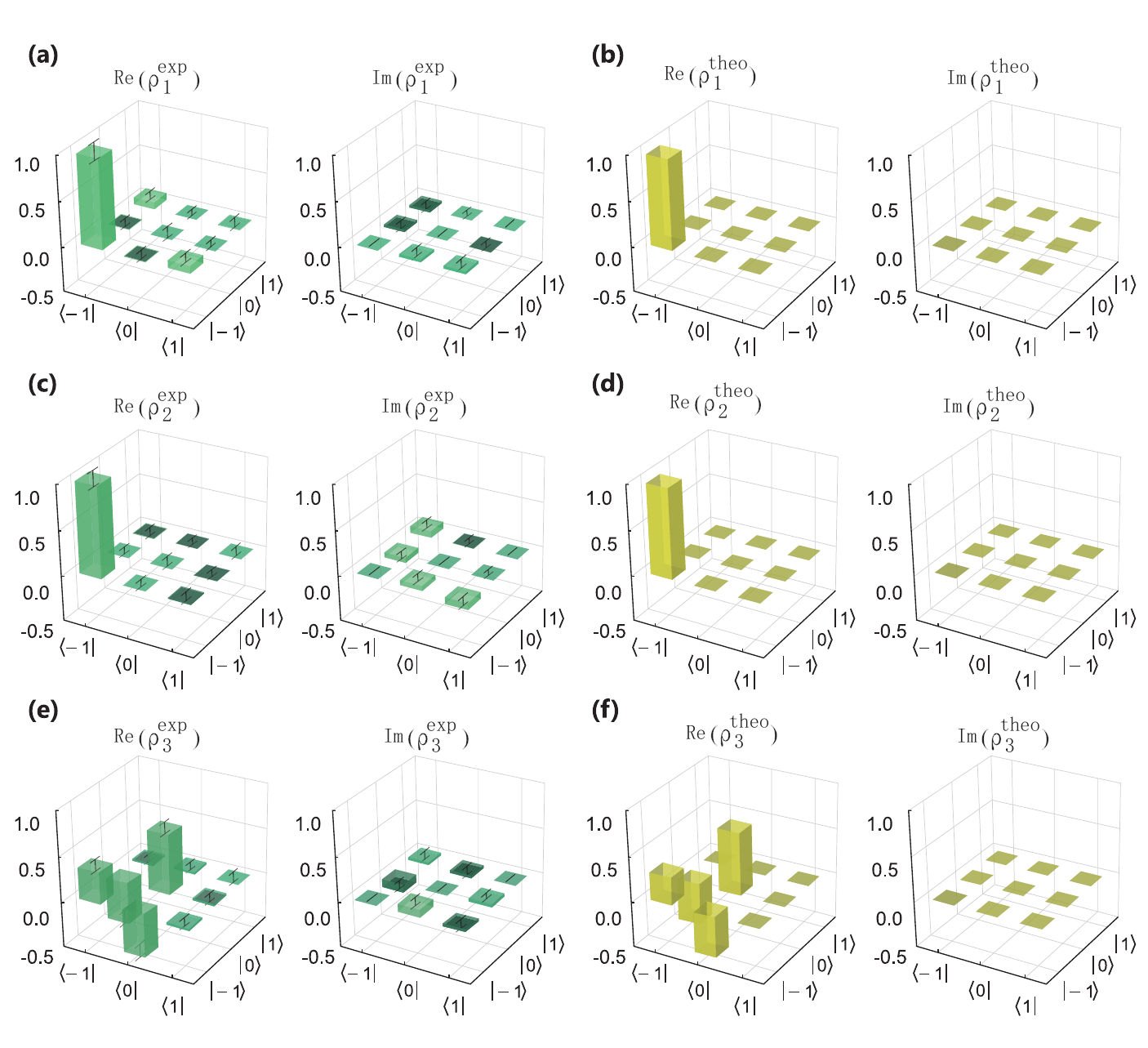}
	\caption{Eigenstates of the non-Hermitian Hamiltonian at the Dirac EP where $k_1=0$ and $k_2=1$. (a,c,e) $\rho_1^{\rm exp}$, $\rho_2^{\rm exp}$ and $\rho_3^{\rm exp}$ are the measured density matrices of three eigenstates (labeled by 1,2 and 3) obtained by quantum state tomography. (b,d,f) $\rho_1^{\rm theo}, \rho_2^{\rm theo}$ and $\rho_3^{\rm theo}$ are the density matrices of theoretically predicted eigenstates.
	}
	\label{Fig4}
\end{figure*}

The degeneracy of two of the eigenstates can distinguish Dirac EP from the Hermitian degeneracy.
The eigenstates of the non-Hermitian Hamiltonian at the Dirac EP are obtained by quantum state tomography as shown in Fig.~\ref{Fig4}.
In order to avoid that the directly obtained state is unphysical, the maximum likelihood estimation method is used to obtain the density matrix.
The overlap between two states $\rho_i$ and $\rho_j$ is characterized by the fidelity, defined as $[\mathrm{Tr}(\sqrt{\sqrt{\rho_i}\rho_j\sqrt{\rho_i}})]^2$.
The obtained fidelities between the experimental eigenstates $\rho_i^{\mathrm{exp}}$ corresponding to $E_i$ and the theoretical ones $\rho_i^{\mathrm{theo}}$ are 0.99(2), 0.99(2) and 0.99(3), yielding the high-fidelity reconstructions of the eigenstates at the Dirac EP.
At the Dirac EP, the experimental results of the fidelity $F_{ij}$ between the experimental eigenstates $\rho_i^{\mathrm{exp}}$ and $\rho_j^{\mathrm{exp}}$ corresponding to the eigenvalues $E_i$ and $E_j$ are $F_{12}=0.98(3)$, $F_{13}=0.32(6)$ and $F_{23}=0.33(5)$.
These results demonstrate the degeneracy of the eigenstates $\rho_1$ and $\rho_2$. 
Both the degeneracy of two eigenstates and corresponding eigenvalues confirms that such a degeneracy is a Dirac EP rather than a Hermitian degeneracy.
\textit{Discussion}- Our work investigates the Dirac EP, a novel type of energy level degeneracies different from typical EPs. 
Around the Dirac EP, the eigenvalues are observed to be real. 
This result reveals that the $\mathcal{PT}$ operator always shares common eigenstates with the non-Hermitian Hamiltonian, and spontaneous $\mathcal{PT}$ symmetry breaking does not occur when passing through the Dirac EP.
The dispersion around the Dirac EP exhibits a conical shape.
The measured degeneracy of the eigenstates further confirms that the Dirac EP is entirely different from Hermitian degeneracy.
The inner product of the eigenstates satisfies a quadratic relationship as the parameters $k_1$ and $k_2$ approach the Dirac EP, which is distinctly different from the linear relationship exhibited in the vicinity of a typically EP (See Supplementary Materials, section 2 for the details).
The occurrence of the Dirac EP is not limited to the model parameters we used, which indicates that similar non-Hermitian degeneracy can potentially be observed in a wider range of systems (See Supplementary Materials, section 2 for the details).

The discovery of the novel type of EP not only provides an unprecedented mechanism for dynamics in non-Hermitian systems but also holds potential application value in the fields of quantum information processing and topological physics.
The Dirac EPs are expected to provide new mechanisms of mode switch that are beyond previous frameworks for typical EPs~\cite{arxiv_Kumar}.
Both chiral and non-chiral mode switch can be realized by passing through the Dirac EP in different directions (See Supplementary Materials, section 4 for the details).
This result unveils a novel mechanism for dynamics in non-Hermitian systems, distinct from the gain and loss arising from the imaginary parts of eigenvalues near typical EPs~\cite{arxiv_Kumar}, which warrants deeper investigation in the future.

The Dirac EPs exhibit advantages compared to typical EPs in realizing mode switch.
Due to its robustness against variations in the pathway, mode switch offers a robust fashion for controlling quantum states, thereby holding potential applications in quantum information processing.
Realizing mode switch by typical EPs suffers from the probability loss due to the imaginary parts of eigenvalues, which makes its application in quantum controls very challenging since the successful post-selection probability is almost zero. 
The real eigenvalues near Dirac EP enable the mode switch to overcome the low efficiency in obtaining the target final state (See Supplementary Materials, section 4 for the details).
Thus, the investigation of Dirac EP provides a novel method to quantum state control and potentially unlocks the applications in quantum information processing.
These results also hold potential in mode switch devices in optical and mechanical systems. 

An application of Dirac EP in the realm of topological physics lies in its facilitation of experimental investigation of complex geometric phases~\cite{PRA_Gong,PRA_Cui,PRA_Zhang}.
Such a complex geometric phase is the generalization of geometric phase, a fundamental concept of topological physics, to the realm of non-Hermitian systems.
Unlike recent work that specifically chooses a trajectory with real spectrum in parameter space~\cite{PRL_Arkhipov}, utilizing Dirac EP offers an alternative approach to achieve adiabatic evolution in non-Hermitian systems, circumventing the non-adiabatic transitions associated with typical EPs.
This enables enabling the investigation of complex geometric phases (See Supplementary Materials, section 3 for the details).

\textit{Acknowledgment}- This work was supported by the Innovation Program for Quantum Science and Technology (Grant No. 2021ZD0302200), the National Natural Science Foundation of China (Grant Nos. T2388102, 12174373 and 12261160569), the Chinese Academy of Sciences (Grant Nos. XDC07000000 and GJJSTD20200001), Hefei Comprehensive National Science Center and the Fundamental Research Funds for the Central Universities (Grant No. WK3540000013).

Yang Wu, Dongfanghao Zhu and Yunhan Wang contributed equally to this work.

\newpage

%
%

\onecolumngrid
\vspace{1.5cm}
\begin{center}
	\textbf{\large Supplementary Material}
\end{center}

\setcounter{figure}{0}
\setcounter{equation}{0}
\setcounter{table}{0}
\makeatletter
\renewcommand{\thefigure}{S\arabic{figure}}
\renewcommand{\theequation}{S\arabic{equation}}
\renewcommand{\thetable}{S\arabic{table}}
\renewcommand{\bibnumfmt}[1]{[RefS#1]}
\renewcommand{\citenumfont}[1]{RefS#1}

\section{S1. Experimental setup}
\begin{figure}[htb]
	\centering
	\includegraphics[width=0.4\linewidth]{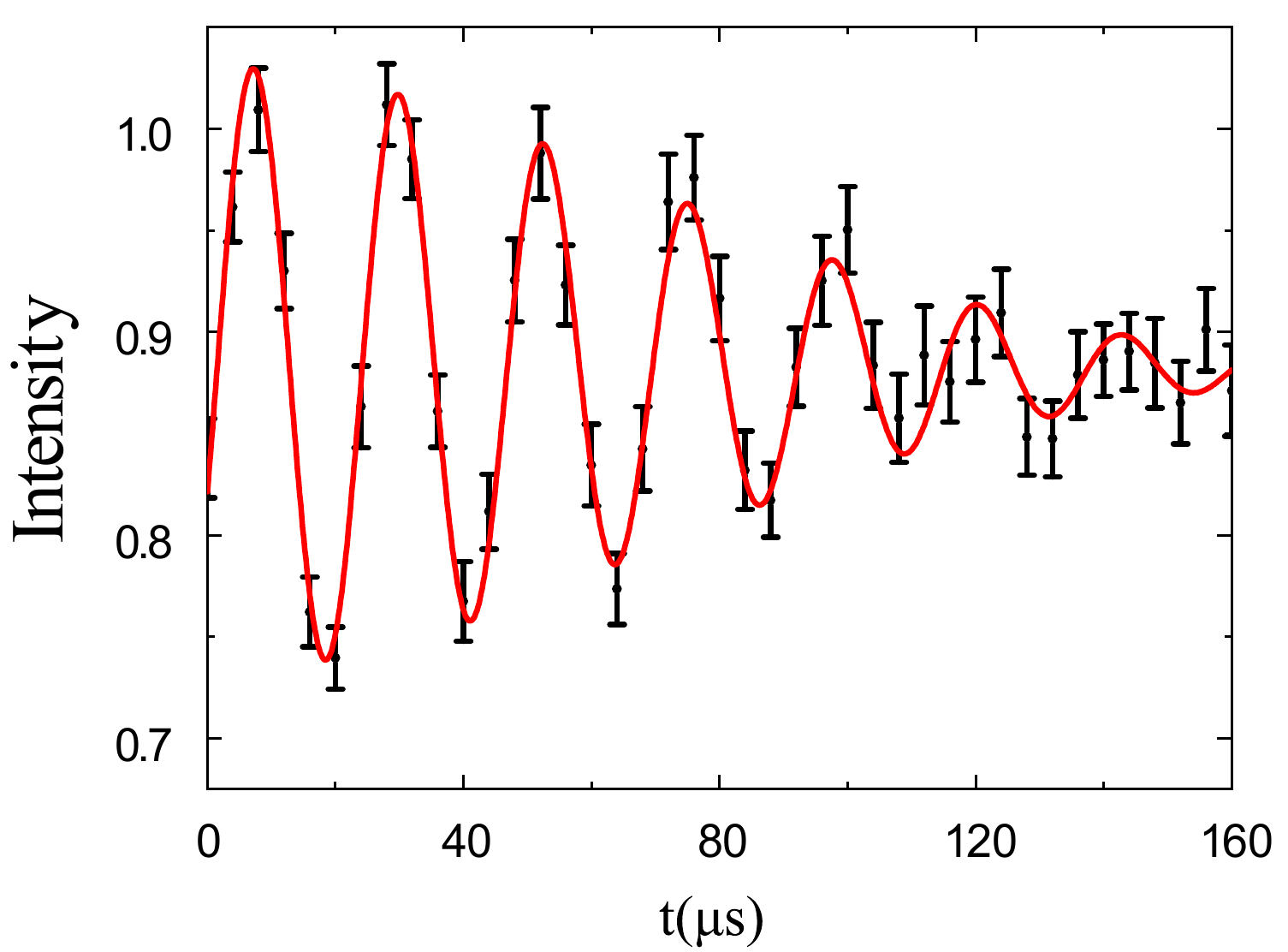}
	\caption{Result of the Ramsey experiment. Fitting the curve gives that $T_2^\star = 96(4)$ $\mathrm{\mu s}$.}
	\label{fig5_1}
\end{figure}

The experiments were implemented on a [100] oriented NV center in an isotopically purified ([$^{12}$C]=99.999\%) diamond synthesized by the chemical vapor deposition method.
The Hamiltonian of the NV center can be written as $H_{\mathrm{NV}} = 2\pi(DS_z^2 + \omega_eS_z + QI_z^2 + \omega_nI_z + AS_zI_z)$,
where $D=2.87$ GHz is the zero-field splitting of the electron spin, $Q=-4.95$ MHz is the nuclear quadrupolar interaction, and $A=-2.16$ MHz is the hyperfine coupling between the electron spin and the nuclear spin.
$\omega_e$ ($\omega_n$) denotes the Zeeman splitting of the electron (nuclear) spin.
$S_z$ and $I_z$ are the spin-1 operators of the electron spin and the nuclear spin, respectively.
The dephasing time of the electron spin of the NV center in our experiments was measured to be $T_2^\star = 96(4)$ $\mathrm{\mu s}$ as shown in Fig.~\ref{fig5_1}.
The diamond was mounted on an optically detected magnetic resonance setup.
The optical excitation was realized by the 532 nm green laser pulses modulated by an acousto-optic modulator (ISOMET).
The laser beam traveled twice through the acousto-optic modulator before going through an oil objective (Olympus, UPLXAPO 100*O, NA 1.45).
The phonon sideband fluorescence (wavelength, 650-800nm) from the NV center went through the same oil objective and was collected by an avalanche photodiode (Perkin Elmer, SPCM-AQRH-14) with a counter card.
The magnetic field of 501 G was provided by a permanent magnet along the NV symmetry axis.
An arbitrary waveform generator (Keysight M8190A) generated MW, electric field and radio-frequency (RF) pulses to manipulate the states of the NV center.
The MW, electric field and RF pulses were amplified individually by power amplifiers (two Mini Circuits ZHL-15W-422-S+ for MW and electric field pulses and LZY-22+ for RF pulses).
The MW pulses and electric fields were applied by transmission line and electrodes fabricated on the surface of diamond, respectively.
The RF pulses were carried by a home-built RF coil.

\section{S2. Non-Hermitian Hamiltonian and the Dirac EP}

According to the tight binding model
\begin{equation}
	H_{\mathrm{tb}}=\sum_{m\in \mathbb{Z}}(m+k_1)^2\ket{m}\bra{m}+\frac{1-k_2}{2}\ket{m}\bra{m+1}+\frac{1+k_2}{2}\ket{m}\bra{m-1},
\end{equation}
the studied Hamiltonian with Dirac EP is the block keeping only the $m=-1, 0$ and $1$ as follows:
\begin{equation}
	\label{Hamiltonian}
	H(k_1,k_2)=\left(\begin{array}{ccc}
		3+2 k_1 & 1-k_2 & 0 \\
		1+k_2 & 0 & 1-k_2 \\
		0 & 1+k_2 & 3-2 k_1
	\end{array}\right),   
\end{equation}
$k_1$ is momentum and $k_2$ characterizes the magnitude of the non-reciprocity of the coupling between adjacent lattices.
The positions of EPs are determined by the the characteristic polynomial $P(E)\equiv \mathrm{det}[H(k_1,k_2)-E]=f_3E^3+f_2E^2+f_1E+f_0$ with multiple roots, where $f_3=-1$, $f_2=6$, $f_1 =4k_1^2-2k_2^2-7$, $f_0 = 6(k_2^2-1)$. 
The polynomial equation $P(E)=0$ has multiple roots when the corresponding discriminant is zero. The discriminant of $P(E)$ takes the form
\begin{equation}
	\Delta(k_1,k_2)=256k_1^6-768k_1^4+2928k_1^2-384k_1^4k_2^2-1824k_1^2k_2^2+192k_1^2k_2^4-168k_2^2+132k_2^4-32k_2^6+68,
\end{equation}
and the Dirac EP appears at $k_1=0, k_2=1$.
For any direction in the two-dimensional space, the eigenvalues satisfy the relationship
\begin{equation}
	E_{1,2}=3+2(-\sin\theta\pm\sqrt{\sin^2\theta+9\cos^2\theta})\Delta k/3+O[(\Delta k)^2],
\end{equation} 
where $\Delta k_1=\cos\theta \Delta k$, $\Delta k_2=\sin\theta \Delta k$, $\theta\in[0,2\pi]$ and $\Delta k>0$. 
Therefore, the linear relationship is satisfied in all directions provided that  the parameters are very close to the Dirac EP, presenting the Dirac cone.
At $k_2=1$, the above linear relationship is transformed into the following form:
\begin{equation}
	\begin{aligned}
		E_1&=3+2|k_1|,\\
		E_2&=3-2|k_1|,\\
		E_3&=0.
	\end{aligned}
\end{equation}
There is a linear relationship between eigenvalues and the parameter $k_1$.
In the direction of $k_2$, for $k_1=0$ and $k_2<1$ the three energy eigenvalues can be represented by $k_2$ as:
\begin{equation}
	\begin{aligned}
		E_1&=\frac{3+\sqrt{17-8k_2{ }^2}}{2},\\
		E_2&=3,\\
		E_3&=\frac{3-\sqrt{17-8k_2{ }^2}}{2},
	\end{aligned}
\end{equation}
and for $k_2>1$
\begin{equation}
	\begin{aligned}
		E_1&=3,\\
		E_2&=\frac{3+\sqrt{17-8k_2{ }^2}}{2},\\
		E_3&=\frac{3-\sqrt{17-8k_2{ }^2}}{2}.
	\end{aligned}
\end{equation}
When the parameter $k_2$ is very close to the Dirac EP at 1, the linear dispersion of the eigenvalues satisfies the following form with $\Delta k_2=k_2-1$
\begin{equation}
	\begin{aligned}
		E_1&=3-4\Delta k_2/3+O[(\Delta k_2)^2],\\
		E_2&=3,
	\end{aligned}
\end{equation}
for the case of $k_2>1$ and
\begin{equation}
	\begin{aligned}
		E_1&=3,\\
		E_2&=3-4\Delta k_2/3+O[(\Delta k_2)^2],
	\end{aligned}
\end{equation}
for the case of  $k_2>1$.
Therefore, in the vicinity of Dirac EP, the eigenvalues at $k_1=0$ also has a linear dependence on $k_2$. 

\begin{figure}[http]
	\centering
	\includegraphics[width=0.6\columnwidth]{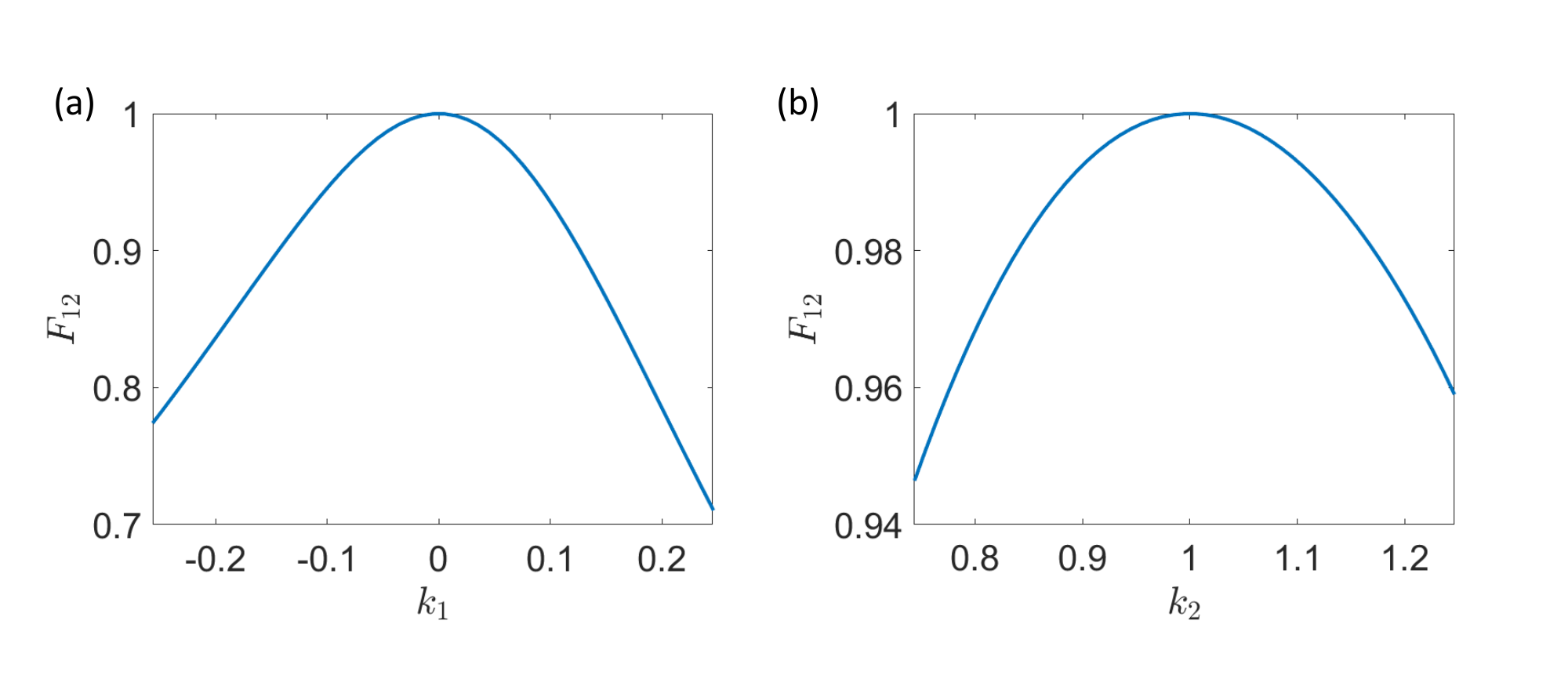}
	\caption{The inner product $F_{12}$ between the two eigenstates of the non-Hermitian Hamiltonian when the parameters $k_1$ (a) and $k_2$ (b) approaching the Dirac EP. 
	}
	\label{overlaps}
\end{figure}

The inner product of the eigenstates as the parameters approach the Dirac EP is also different from the typical EP.
In the directions of parameter changes for $k_1$ and $k_2$ approaching the Dirac EP, the inner product, $F_{12}=|\braket{\psi_1|\psi_2}|$, between the two eigenstates satisfies the relations $F_{12}\propto(\Delta k_1)^2$ and $F_{12}\propto(\Delta k_2)^2$ as shown in the Fig.~\ref{overlaps}. This is distinctly different from the linear relation exhibited at a typical EP.

\begin{figure}[http]
	\centering
	\includegraphics[width=0.4\columnwidth]{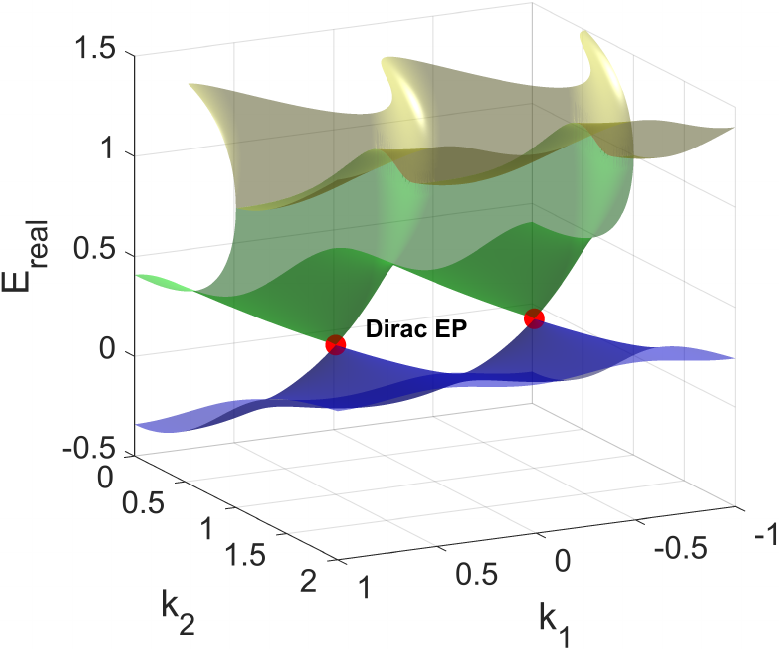}
	\caption{The eigenvalues structure and the Dirac exceptional points of the non-Hermitian Hamiltonian with next-nearest-neighbor coupling. The Dirac EPs marked by the red dots are located at $k_1=\pm0.5$. 
	}
	\label{new_point}
\end{figure}

The occurrence of the Dirac EP is not limited to the model parameters we used. 
Considering the existence of next-nearest-neighbor coupling, the tight-binding model is 
\begin{equation}
	H_{\mathrm{tb,nn}}=\sum_{m\in \mathbb{Z}}(m+k_1)^2\ket{m}\bra{m}+\frac{1-k_2}{2}\ket{m}\bra{m+1}+\frac{1+k_2}{2}\ket{m}\bra{m-1}+\frac{1-k_2}{2}\ket{m}\bra{m+2}+\frac{1+k_2}{2}\ket{m}\bra{m-2}.
\end{equation}
When keeping only the $m = 0, \pm1$, and $\pm2$ block, the eigenvalues structure of the non-Hermitian Hamiltonian is shown in Fig.~\ref{new_point}.
We can found that the Dirac EP are located at $k_1=\pm0.5$. 
Therefore, besides the case of $k_1=0$, Dirac EPs can be constructed under other parameters as well.

	\section{S3. The geometric phase around the Dirac EP}
	The geometric phase, as a fundamental concept in topological physics, has been extended to non-Hermitian systems\cite{PRA_Cui,PRA_Zhang}.
	The complex eigenvalues near the typical EP leads to the inevitable non-adiabatic transitions in the evolution\cite{JPMT_Uzdin,NRP_Ding}, making experimental investigation on EP-related geometric phases challenging. 
	The real energy spectrum near the Dirac EP enables adiabatic evolution. 
	This property makes it possible to experimentally observe the complex geometric phase around EP, which is absent in Hermitian systems.
	Here, we give an example as follows.
	For the model Hamiltonian, we implement the encircling Dirac EP by changing the parameters $k_1$ and $k_2$ in time according to the relationship $k_1(t)=R\cos(\omega t)$ and $k_2(t)=1-R\sin(\omega t)$, where $R$ is the encircling radius and $\omega$ is the angular velocity of encircling.
	$R$ is set to 0.2 to ensure that the eigenvalues of the Hamiltonian on the path are all real.
	$\omega$ is set to $2\pi/T$, where $T=5000$ to ensure that the evolution is always adiabatic.
	The total phase accumulated during the evolution is
	\begin{equation}
		\phi_{\mathrm{tot}}=i\ln(\bra{\psi_i}\mathcal{T}e^{-i\int H(k_1(t),k_2(t))dt}\ket{\psi_i}),
	\end{equation}
	where $\ket{\psi_i}$ is the initial state, which is the eigenstate of the non-Hermitian Hamiltonian $H(k_1(t),k_2(t))$ at $t=0$ corresponding to eigenvalue $E_i$.
	The geometric phase $\phi_{\mathrm{geo}}$ can be obtained by
	\begin{equation}
		\phi_{\mathrm{geo}}=\phi_{\mathrm{tot}}-\phi_{\mathrm{dyn}}, 
	\end{equation}
	where $\phi_{\mathrm{dyn}}$ is the dynamical phase defined as $\phi_{\mathrm{dyn}} = \int E_i(t)dt$.
	According to the parameters we set, the geometric phases for the three initial eigenstates are $-0.14i$, $0.1i$ and $0.04i$ respectively.
	Furthermore, we changed the angular velocity of encircling while ensuring adiabaticity and found that the imaginary geometric phase did not change, which shows that it does reflect the geometric properties of the evolution path.
	
	\section{S4. Mode switch by passing through the Dirac EP}
\begin{figure}[http]
		\centering
		\includegraphics[width=0.7\columnwidth]{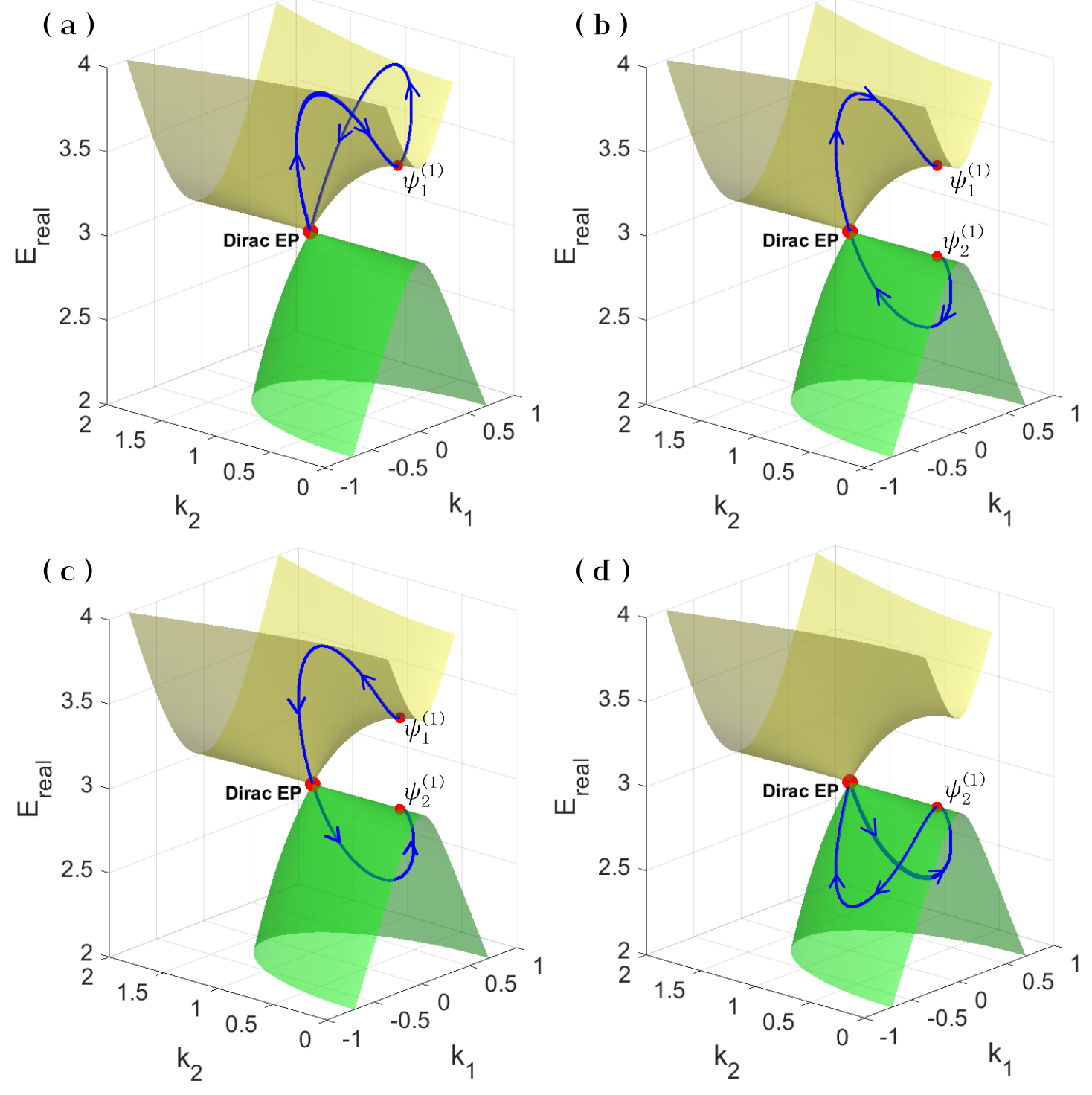}
		\caption{Chiral mode switch by passing through the Dirac EP from the $k_2$ direction. (a-d) The prospective trajectories in the eigenvalue sheets when the direction is counterclockwise (a,b) and clockwise (c,d). The initial state is $\ket{\psi_1^{(1)}}$ (a,c) and $\ket{\psi_2^{(1)}}$ (b,d).
		}
		\label{switch1}
	\end{figure}

\begin{figure}[http]
		\centering
		\includegraphics[width=0.7\columnwidth]{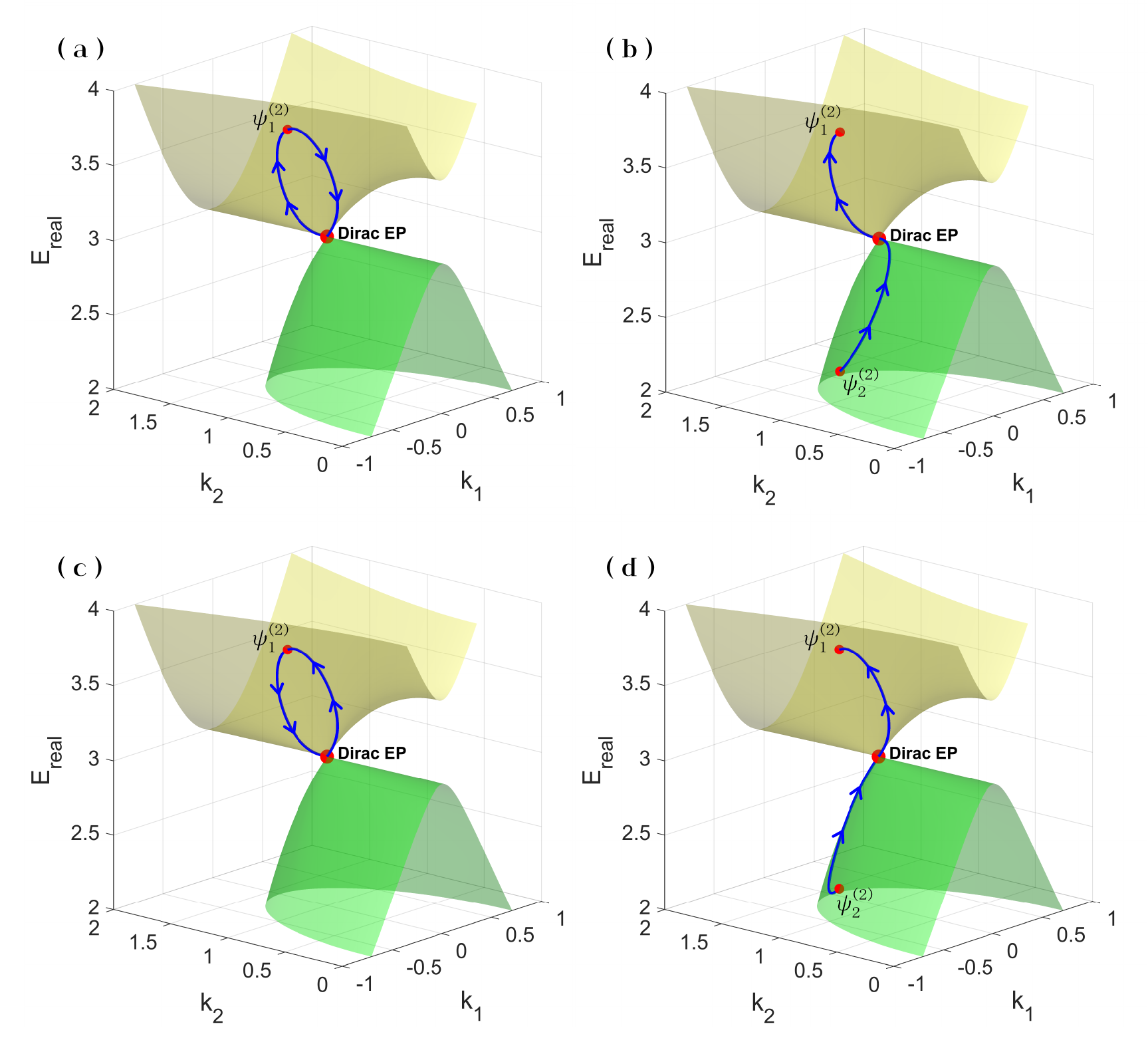}
		\caption{Non-chiral mode switch by passing through the Dirac EP from the $k_1$ direction. (a-d) The prospective trajectories in the eigenvalue sheets when the direction is counterclockwise (a,b) and clockwise (c,d). The initial state is $\ket{\psi_1^{(2)}}$ (a,c) and $\ket{\psi_2^{(2)}}$ (b,d).
		}
		\label{switch2}
	\end{figure}

	Mode switch is an exotic phenomenon related to EP in non-Hermitian systems.
	Due to the dissipation introduced by the complex eigenvalues near the typical EP, mode switch suffers from the low efficiency in obtaining the target final state\cite{PRX_Zhang,PRL_Liu}. 
	The real eigenvalues near Dirac EP and the degeneracy of eigenstates at  Dirac EP provide new mechanisms of mode switch that are beyond previous frameworks for typical EPs.
	Both chiral and non-chiral mode switch can be realized by passing through the Dirac EP in the different directions.
	Here we give examples as follows.
	Firstly, we show that the chiral mode switch is realized by passing through the Dirac EP from the $k_2$ direction.
	The parameters $k_1$ and $k_2$ are set to change with time according to the relationship $k_1(t)=0.4\sin(\omega_s t)$ and $k_2(t)=0.6-0.4\cos(\omega_s t)$. 
	This setting can ensure that the path passes through Dirac EP from the $k_2$ direction.
	The angular velocity $\omega_s$ is set to $2\pi/T_s$, where $T_s=1~\mathrm{s}$ to ensure that the evolution is adiabatic except when passing through Dirac EP.
	The initial state is prepare to the eigenstates $\ket{\psi_1^{(1)}}$ (Fig.~\ref{switch1}(a,c)) and $\ket{\psi_2^{(1)}}$ (Fig.~\ref{switch1}(b,d)) of non-Hermitian Hamiltonian at $t=0$. 
	When the pathway is counterclockwise, both $\ket{\psi_1^{(1)}}$ and $\ket{\psi_2^{(1)}}$ will evolve to $\ket{\psi_1^{(1)}}$ (Fig.~\ref{switch1}(a,b)).
	When the direction is clockwise, both $\ket{\psi_1^{(1)}}$ and $\ket{\psi_2^{(1)}}$ will evolve to $\ket{\psi_2^{(1)}}$ (Fig.~\ref{switch1}(c,d)).
	The final state depends on the path direction, which is taken as the chiral mode switch.
	Secondly, we show that the non-chiral mode switch is realized by passing through the Dirac EP from the $k_1$ direction.
	The parameters $k_1$ and $k_2$ are changed with time corresponding to the relationship $k_1(t)=-0.2-0.2\cos(\omega_s t)$ and $k_2(t)=1+0.2\sin(\omega_s t)$.
	The angular velocity $\omega_s$ is also set to $2\pi/T_s$ with $T_s=1~\mathrm{s}$.
	Regardless of whether the initial state is $\ket{\psi_1^{(2)}}$ (Fig.~\ref{switch2}(a,c)) and $\ket{\psi_2^{(2)}}$ (Fig.~\ref{switch2}(b,d)) and the pathway is counterclockwise (Fig.~\ref{switch2}(a,b)) or clockwise (Fig.~\ref{switch2}(c,d)), the final state is always $\ket{\psi_1^{(2)}}$.
	The results of passing through the Dirac EP from the $k_1$ direction show the non-chiral mode switch.
	In the cases we show, the overlaps between the final state of the evolution and the target state are higher than 0.99, and the transmission efficiencies are higher than 60\%.
	
	\section{S5. The dilation method and the experimental realization in NV center system}
	
	This section introduces the details of realizing the dilated Hamiltonian in NV center. 
	To realize the NH Hamiltonian in Eq.~\eqref{Hamiltonian}, we introduce an ancilla qubit and apply the universal dilation method to obtain an Hermitian Hamiltonian $H_{\mathrm{tot}}$. 
	The dilated Hamiltonian can be expanded as:
	\begin{equation}
		H_{\mathrm{tot}}(t)= \Gamma(t) \otimes|1\rangle_{n n}\langle 1| +\Lambda(t) \otimes| 0\rangle_{n n}\langle 0|.
	\end{equation}
	Here $\Lambda(t)$ and $\Gamma({t})$ are determined on the NH Hamiltonian $H$ from $\Gamma(t)=\hat{\Lambda}(t)+\hat{\Gamma}(t)$ and $\Lambda(t)=\hat{\Lambda}(t)-\hat{\Gamma}(t)$
	\begin{align}\footnotesize
		&\hat{\Lambda}(t)=\left\{H(t)+\left[i \frac{\mathrm{d}}{\mathrm{d}t} \eta(t)+\eta(t) H(t)\right] \eta(t)\right\} M^{-1}(t), \\
		&\hat{\Gamma}(t)=i\left[H(t) \eta(t)-\eta(t) H(t)-i \frac{\mathrm{d}}{\mathrm{d}t} \eta(t)\right] M^{-1}(t).
	\end{align}
	The explicit forms of $M$ and $\eta$ are 
	\begin{equation}
		M(t) = e^{-iH^\dagger t}M(0)e^{iHt},\eta(t)=U(t)[M(t)-I]^{1/2}.
	\end{equation}
	The choice of $M(0)$ and $U(t)$ is flexible. 
	The details of the derivations can be found in Ref.\cite{Dilation}. 
	In our experiment, $M(0)=1.3I$ is chosen. 
	In order to facilitate the realization of $H_{\rm tot}$ in NV center, $\hat{H}_{\rm tot} = s H_{\rm tot}$ is implemented, where $s$ is a nonzero coefficient that scales the evolution time. This is equivalent to apply the dilation method to the NH Hamiltonian $sH$ instead of $H$ (in our experiment, $s$ is of the order of 30-50 kHz, depending on different points). 
	
\begin{figure}[http]
		\centering
		\includegraphics[width=0.8\columnwidth]{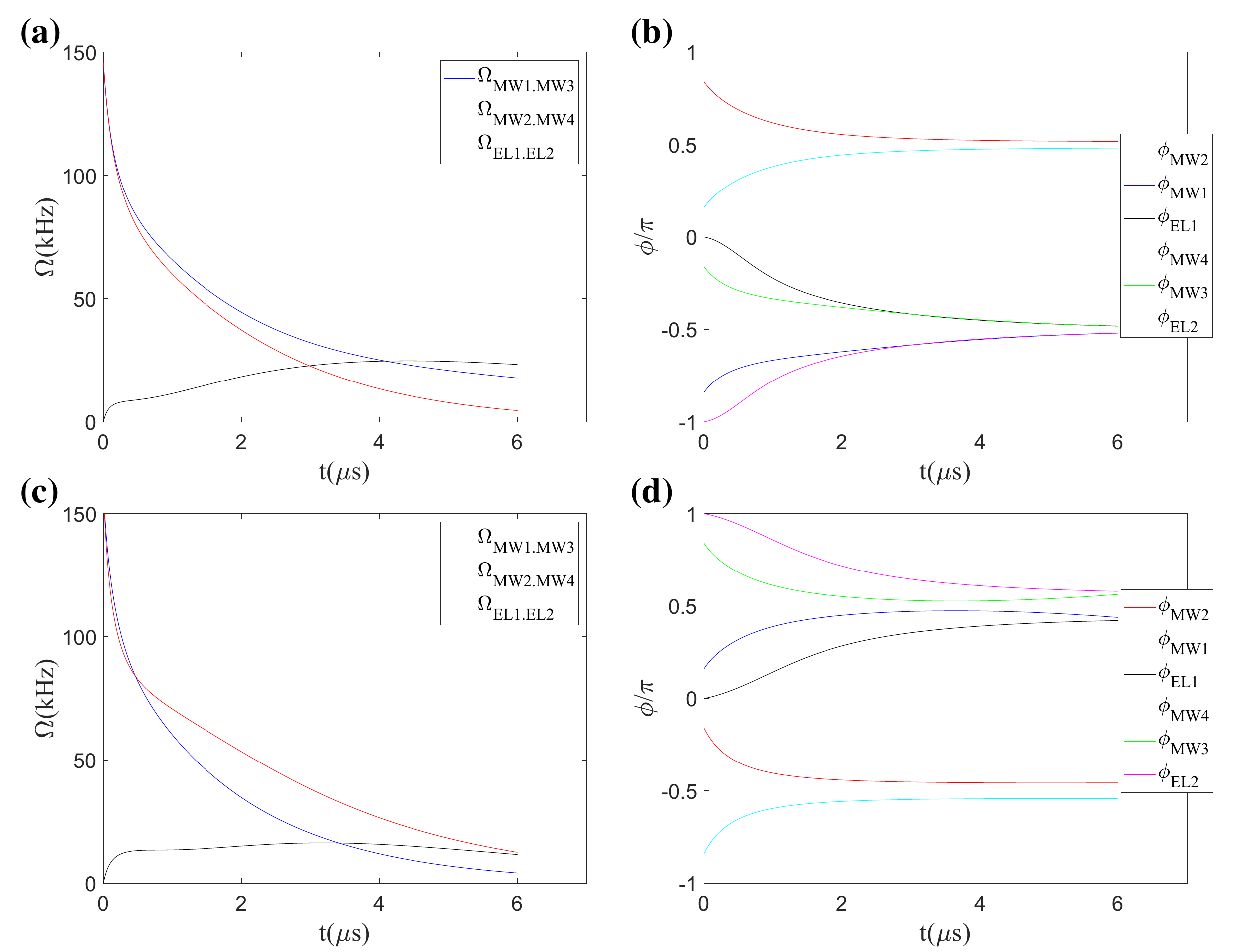}
		\caption{The amplitudes ($\Omega_{\alpha}$) (a,c) and phases ($\phi_{\alpha}$) (b,d) of the microwave and electric field pulses for the parameter configurations $k_1=0$, $k_2=1$ (a,b) and $k_1=0$, $k_2=1.3$ (c,d).
		}
		\label{amp_phase}
	\end{figure}

	Then we focus on the design of pulse parameters for implementing $H_{\rm tot}$ in NV center. $H_{\rm tot}$ is realized in NV center by implementing a control Hamiltonian and choosing an appropriate interaction picture. The control Hamiltonian takes the following form (the integral variable $\tau$ is omitted)
	\begin{equation}
		\begin{aligned}
			H_C(t) &= 2\sqrt{2}\pi\Omega_{\rm MW2} \cos[\int_{0}^{t}\omega_{\rm MW2}+\phi_{\rm MW2}(t)]S_x\otimes\ket{1}\bra{1}\\
			&+2\sqrt{2}\pi\Omega_{\rm MW1}  \cos[\int_{0}^{t}\omega_{\rm MW1}+\phi_{\rm MW1}(t)]S_x\otimes\ket{1}\bra{1}\\
			&+2\pi\Omega_{\rm EF1}  \cos[\int_{0}^{t}\omega_{\rm EF1}+\phi_{\rm EF1}(t)]S_{13}\otimes\ket{1}\bra{1}\\
			&+2\sqrt{2}\pi\Omega_{\rm MW4}  \cos[\int_{0}^{t}\omega_{\rm MW4}+\phi_{\rm MW4}(t)]S_x\otimes\ket{0}\bra{0}\\
			&+2\sqrt{2}\pi\Omega_{\rm MW3}  \cos[\int_{0}^{t}\omega_{\rm MW3}+\phi_{\rm MW3}(t)]S_x\otimes\ket{0}\bra{0}\\
			&+2\pi\Omega_{\rm EF2}  \cos[\int_{0}^{t}\omega_{\rm EF2}+\phi_{\rm EF2}(t)]S_{13}\otimes\ket{0}\bra{0},
		\end{aligned}
	\end{equation}
	where 
	\begin{equation}
		S_x = \frac{1}{\sqrt{2}}\left[\begin{array}{ccc}
			0 & 1 & 0\\
			1 & 0 & 1\\
			0 & 1 & 0
		\end{array}\right], S_{13} = \left[\begin{array}{ccc}
			0 & 0 & 1\\
			0 & 0 & 0\\
			1 & 0 & 0
		\end{array}\right].
	\end{equation}
	The parameters to be determined are $\Omega_{\zeta}$, $\phi_{\zeta}$ and $\omega_{\zeta}$, where $\zeta\in\lbrace\rm MW1,MW2,MW3,MW4,EF1,EF2\rbrace$.
	The interaction picture is chosen as
	\begin{equation}
		U_{\rm rot} = \exp[i\int_{0}^{t}H_{\rm NV}-\text{diag}(d_1,d_2,\dots,d_6)],
	\end{equation}
	where $H_{\rm NV}$ is the Hamiltonian for the ground state of NV center and $d_1,d_2,\dots,d_6$ are the diagonal elements of $H_{\rm tot}$.
	The Hamiltonian under the rotating wave approximation can be written as
	\begin{equation}
		\begin{aligned}
			&H_{\rm rot}=U_{\rm rot}H_CU_{\rm rot}^\dagger+\text{diag}(d_1,d_2,\dots,d_6)\\
			&=\left[\begin{array}{ccc}
				d_1 & A_1 & C_1\\
				A_1^* & d_2 & B_1\\
				C_1^* & B_1^* & d_3
			\end{array}\right]\otimes\ket{1}\bra{1}+\left[\begin{array}{ccc}
				d_4 & A_2 & C_2\\
				A_2^* & d_5 & B_2\\
				C_2^* & B_2^* & d_6
			\end{array}\right]\otimes\ket{0}\bra{0},
		\end{aligned}
	\end{equation}
	where
	\begin{equation}
		\begin{aligned}
			&A_1=\pi\Omega_{\rm MW2} e^{-i\phi_{\rm MW2}-i(\int_{0}^{t}\omega_{\rm MW2}-\omega_{12}-d_2+d_1)},\\
			&B_1=\pi\Omega_{\rm MW1} e^{i\phi_{\rm MW1}+i(\int_{0}^{t}\omega_{\rm MW1}-\omega_{23}-d_2+d_3)},\\
			&C_1=\pi\Omega_{\rm EF1} e^{-i\phi_{\rm EF1}-i(\int_{0}^{t}\omega_{\rm EF1}-\omega_{13}-d_3+d_1)},\\
			&A_2=\pi\Omega_{\rm MW4} e^{-i\phi_{\rm MW4}-i(\int_{0}^{t}\omega_{\rm MW4}-\omega_{45}-d_5+d_4)},\\
			&B_2=\pi\Omega_{\rm MW3} e^{i\phi_{\rm MW3}+i(\int_{0}^{t}\omega_{\rm MW3}-\omega_{56}-d_5+d_6)},\\
			&C_2=\pi\Omega_{\rm EF2} e^{-i\phi_{\rm EF2}-i(\int_{0}^{t}\omega_{\rm EF2}-\omega_{46}-d_6+d_4)}.\\
		\end{aligned}
	\end{equation}
	Here we label the energy levels $\ket{m_S=1,0,-1}_e\otimes\ket{m_I=1}_n$ ($\ket{m_S=1,0,-1}_e\otimes\ket{m_I=0}_n$) as 1, 2 and 3 (4, 5 and 6) for simplicity and $\omega_{ij}$ $(\omega_{ij}>0)$ is the transition frequency between the levels $i$ and $j$.
	Comparing $H_{\rm rot}$ and $H_{\rm tot}$, all the parameters $\Omega_{\zeta}$, $\phi_{\zeta}$ and $\omega_{\zeta}$ can be solved.
	Two examples of the amplitudes and phases of the microwave and electric field pulses are shown in 
	Fig.~\ref{amp_phase}a,c and Fig.~\ref{amp_phase}b,d for the parameter configurations $k_1=0$, $k_2=1$ and $k_1=0$, $k_2=1.3$, respectively.
	
\begin{figure}[http]
		\centering
		\includegraphics[width=0.8\columnwidth]{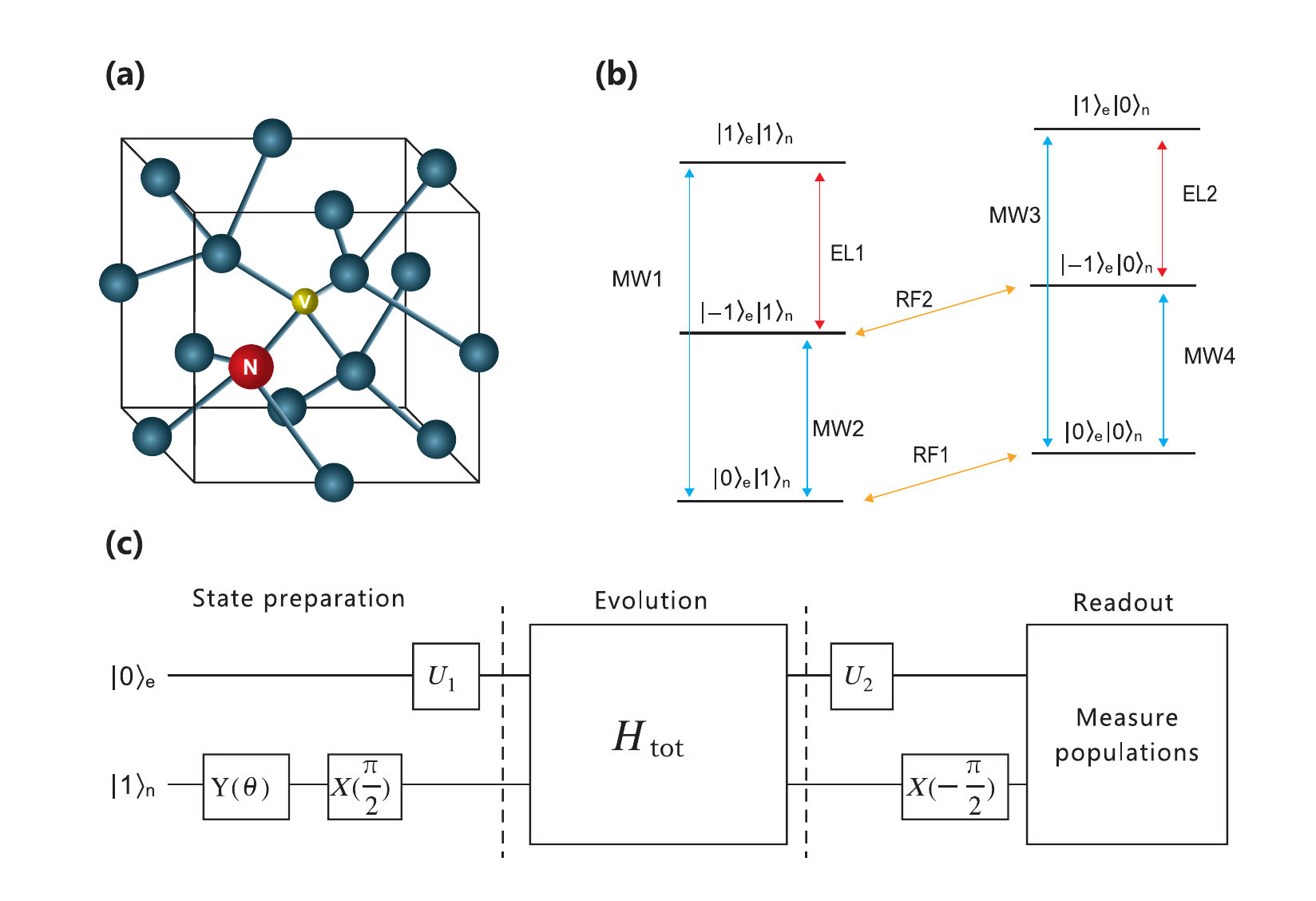}
		\caption{Constructing of the non-Hermitian Hamiltonian in an NV center in diamond. Schematic representation of the atomic structure for the NV center, comprising by a substitution nitrogen atom (red ball) adjacent to a vacancy (yellow ball). (b) The coupled system of the electron and nuclear spin in NV center to construct the non-Hermitian Hamiltonian is composed of six energy levels $\ket{1, 0, -1}_e\ket{1}_n$ and $\ket{1, 0, -1}_e\ket{0}_n$, where the subscripts $e$ and $n$ correspond to the electron spin and nuclear spin, respectively. MW (blue arrows) and EL pulses (red arrows)	are implemented to realize the non-Hermitian Hamiltonian and manipulate the electron spin state. The RF pulses (orange arrows) are applied for the initial state preparation and the measurement	of the nuclear spin. (c) Quantum circuit of the experiment. X and Y denote the single nuclear	spin qubit rotation around the $x$ and $y$ axes. The initial state of the coupled system is prepared by rotations Y($\theta$) and X($\pi/2$) combined with the operation $U_1$ on the electron spin. Then the coupled	system evolves under the dilation Hamiltonian $H_{tot}(t)$. The populations of the six energy levels are measured after the rotation X($-\pi/2$) and the operation $U_2$.
		}
		\label{Circuit}
	\end{figure}

	The nitrogen-vacancy (NV) center, which is an atomic-scale defect in diamond as shown in Fig.~\ref{Circuit}a, was utilized to observe the Dirac EP.
	Taking the electron spin in NV center as the system and the nuclear spin as the ancilla qubit, the non-Hermitian Hamiltonian can be realized based on the dilation method. 
	The subspace spanned by $|m_S\rangle_e|1\rangle_n$ and $|m_S\rangle_e|0\rangle_n$ with $m_S=1, 0, -1$ was utilized to construct the dilated Hamiltonian corresponding to the non-Hermitian Hamiltonian. 
	Here subscripts $e$ and $n$ label the electron and nuclear spin states, respectively.
	The dilated Hamiltonian is realized by four selective microwave pulses (blue arrows in Fig.~\ref{Circuit}b) and two alternating current electric field pulses (red arrows in Fig.~\ref{Circuit}b).
	The amplitudes and phases of the microwave and electric field pulses were appropriately set according to the off-diagonal elements in $H_{\rm tot}(t)$.
	%
	%
	The circuit of the experiment is shown in Fig.~\ref{Circuit}c.
	The initial state is polarized to $|0\rangle_e|1\rangle_n$ by optical pumping under the static magnetic field of 501 Gauss.
	After the polarization, the initial state of the coupled system is prepared to the form $|\psi\rangle_e\otimes(|-\rangle_n+\eta_0|+\rangle_n)$. 
	Here, $\eta_0$ is properly chosen for the convenience of experimental realization, and $|\pm\rangle_n$ are the eigenstates of the Pauli matrix $\sigma_y$.
	The initial state of the nuclear spin was prepared by the rotation $Y(\theta)$ followed by the rotation $X(\pi/2)$ of the nuclear spin.
	The operator, $Y(\theta)$, stands for the rotation around \textit y axis with the rotation angle $\theta = 2\arctan(\eta_0)$.
	The rotation $X(\pi/2)$ is the single-qubit rotations around \textit x axis to realize transformation between the basis spanned by $\{|0\rangle_n, |1\rangle_n\}$ and $\{|+\rangle_n, |-\rangle_n\}$ of the nuclear spin qubit.
	The nuclear spin rotations were realized by radio-frequency (RF) pulses (orange arrows in Fig.~\ref{Circuit}b). 
	By selecting an appropriate $U_1$, the electronic spin state is prepared to $\ket{\psi}$, ensuring a sufficiently high population in the $\ket{-}$ subspace during readout.
	Then the system evolved under the dilated Hamiltonian $H_{\mathrm{tot}}(t)$.
	Finally, the basis of the electron spin is transformed by applying the rotation of $U_2$.
	The rotation $X(-\pi/2)$ transforms the basis of the nuclear spin from $\{|+\rangle_n, |-\rangle_n\}$ to $\{|0\rangle_n, |1\rangle_n\}$, thereby enabling measurements of populations and quantum state tomography.
	
	\section{S6. Acquisition of the eigenstates and eigenvalues of the NH Hamiltonian}
	
	The eigenstates of an NH Hamitonian $H$ can be obtained from the steady states under the evolution of the NH Hamiltonian $g(H)$, where $g(x)$ is an analytic function of $x$. We label the eigenvalues of $g(H)$ as $\epsilon_{1,2,3}=g(E_{1,2,3})$ with the corresponding eigenstates as $\ket{\psi_{1,2,3}}$. Then for any initial state $\ket{\psi_{\rm ini}} = c_1\ket{\psi_1}+c_2\ket{\psi_2}+c_3\ket{\psi_3}$ ($c_1c_2c_3\ne0$), the evolution governed by $g(H)$ gives
	\begin{equation}
		\ket{\psi} = c_1e^{-i\epsilon_{1}t}\ket{\psi_1}+c_2e^{-i\epsilon_{2}t}\ket{\psi_2}+c_3e^{-i\epsilon_{3}t}\ket{\psi_3}.
	\end{equation}
	Thus when the evolution time $t$ is long enough, the state approaches the eigenstate corresponding to the eigenvalue with largest imaginary part. By changing the form of $g(H)$, all three eigenstates can be approached. With the help of dilation method, as well as the quantum state tomography applied on the subspace where the nuclear spin is $\ket{-}_n$, the density matrix of the eigenstates can be obtained. Since direct results from quantum state tomography may give an unphysical state, the maximum likelihood estimation method was employed to obtain physical results. For the model Hamiltonian, we can choose the form of $g(H)$ from $iH$, $-iH$ and $1/i(H-cI)$ to obtain desired eigenstates, where $c$ is a properly chosen complex number which ensures that the eigenstate corresponding the eigenvalue closest to $c$ can be obtained. 
	
	The eigenvalues are solved from a set of equations according to the model. Here in this section, the details will be shown as follows. 
	Since the characteristic polynomial can also be written as $P(E) = (E-E_1)(E-E_2)(E-E_3)$, we have 
	\begin{equation}
		\label{equ:2}
		\begin{aligned}
			E_1+E_2+E_3&=6,\\
			E_1E_2+E_2E_3+E_1E_3&=7-4k_1{ }^2+2k_2{ }^2,\\
			E_1E_2E_3&=6k_2{ }^2-6,
		\end{aligned}
	\end{equation}
	where $E_i$ is the eigenvalues of the Hamiltonian.
	Thus the eigenvalues can be regarded as variables of the Hamiltonian (instead of the parameters $k_1,k_2$) with the constraint equation
	\begin{equation}
		\label{constraint equation}
		E_1+E_2+E_3=6.
	\end{equation}
	It can be shown that the eigenstate corresponding to the eigenvalue $E_i$ takes the form (not normalized for convenience)
	\begin{equation}\label{equ:states}
		\left|\psi_i\right\rangle=\left(\begin{array}{c}
			E_i\left(E_i+2 k_1-3\right)+k_2{ }^2-1 \\
			\left(E_i+2 k_1-3\right)\left(1+k_2\right) \\
			\left(1+k_2\right)^2
		\end{array}\right).   
	\end{equation}
	According to Eq.\ref{equ:2}, we can replace $k_1,k_2$ by $E_i$ as 
	\begin{equation}
		\label{equ:3}
		\begin{aligned}
			12k_1^2&=27-3(E_1E_2+E_2E_3+E_1E_3)+E_1E_2E_3,\\
			6k_2^2&=E_1E_2E_3+6.
		\end{aligned}
	\end{equation}
	The eigenstates are then parameterized by the eigenvalues only. 
	The other equations can be formulated to relate the three eigenvalues to population information of the eigenstates, which can be obtained through performing population measurements. Define the unitary operations as
	\begin{equation}
		\begin{aligned}
			&U(a,b) = \left[\begin{array}{ccc}
				\cos a & -i \sin (a) e^{-i b} & 0 \\
				-{i} \sin (a) e^{i b} & \cos a & 0 \\
				0 & 0 & 1
			\end{array}\right],\\
			&V(c,d) = \left[\begin{array}{ccc}
				1 & 0 & 0 \\
				0 & \cos c & -{i} \sin (c) e^{-i d} \\
				0 & -{i} \sin (c) e^{i d} & \cos c
			\end{array}\right].\\
		\end{aligned}
	\end{equation}
	These operations can all be realized by microwave pulses. 
	The operations $U_1$ and $U_2$ in Fig. 2c is realized by $U_1=U(a_1,b_1)V(c_1,d_1)$ and $U_2=U(a_2,b_2)V(c_2,d_2)$.
	Denote $P^{(i)}_j$ as the populations measured after the unitary operations $U_2^{(i)}=U(a_2^{(i)},b_2^{(i)})V(c_2^{(i)},d_2^{(i)})$ applied on the three eigenstate $\ket{\psi_i}$, respectively. 
	Here $i = 1,2,3$ label the eigenstates corresponding to the eigenvalues $E_i$, and $j = 1,2,3$ label the populations on the three levels $\ket{1}$, $\ket{0}$ and $\ket{-1}$. 
	We choose the quantities to be measured as $P^{(1)}_2/P^{(1)}_3$, $P^{(2)}_2/P^{(2)}_3$ and $P^{(3)}_2/P^{(3)}_3$.
	The explicit form of the equations can be formulated as
	\begin{equation}
		\left\{\begin{array}{l}
			F^{(1)}(E_1,E_2,E_3)=\frac{P^{(1)}_2}{P^{(1)}_3} \\
			F^{(2)}(E_1,E_2,E_3)=\frac{P^{(2)}_2}{P^{(2)}_3} \\
			F^{(3)}(E_1,E_2,E_3)=\frac{P^{(3)}_2}{P^{(3)}_3}
		\end{array}\right.,
		\label{eqmeas}
	\end{equation}
	where analytical expression of the eigenvalues is 
	\begin{equation}
		F^{(i)}(E_1,E_2,E_3)=\left|\frac{\langle0|U_2^{(i)}|\psi_i\rangle}{\langle-1|U_2^{(i)}|\psi_i\rangle}\right|^2.
	\end{equation}
	Combining Eq.\ref{equ:2} and Eq.\ref{eqmeas}, the three eigenvalues can be solved.



\end{document}